# THE EIC SCIENCE CASE

R.G. Milner[#], Laboratory for Nuclear Science, MIT, Cambridge, MA 02139, USA

*Abstract*

For the first time, physicists are in the position to precisely study a fully relativistic quantum field theory: Quantum ChromoDynamics (QCD). QCD is a central element of the Standard Model and provides the theoretical framework for understanding the strong interaction. This demands a powerful new electron microscope to probe the virtual particles of QCD. *Ab initio* calculations using lattice gauge theory on the world's most powerful supercomputers are essential for comparison with the data. The new accelerator and computing techniques demand aggressive development of challenging, innovative technologies.

## SCIENTIFIC MOTIVATION

Physicists strive to describe the physical universe in the simplest, most fundamental terms that are consistent with experiment. A key element in advancing understanding is to precisely compare theory with experiment, particularly in previously unexplored regimes.

The Standard Model of the subatomic world plus General Relativity successfully describe all observations to date made directly by physicists on the luminous matter in the universe. In the Standard Model, the exchange of gauge bosons mediates the strong, weak, and electromagnetic fundamental interactions between point-like fermions, the quarks and leptons. The electroweak theory unifies the electromagnetic and weak interactions and has been validated most recently by the discovery of the Higgs boson in 2012 at the LHC. This explains how the quarks and charged leptons acquire mass. Astrophysical observations on the large-scale structures in the universe have been interpreted as evidence of a much larger amount of non-luminous dark matter, which is unexplained by the Standard Model, and which is widely sought in experiments.

The strong interaction, which is responsible for the binding of atomic nuclei, drives energy generation in stars and is the basis for the origin of the chemical elements. Physicists have developed a successful description of the strong force in terms of hadrons, which explains nuclear properties and reactions at low energies. A major research thrust at present is the study of the limits of nuclear stability and the new FRIB accelerator will provide world-leading capabilities in this area.

Quantum ChromoDynamics (QCD) is the gauge theory of the strong interaction in the Standard Model. This theory not only explains the structure of hadrons but must provide the fundamental framework to understand the properties and structure of atomic nuclei. In QCD, the six quarks acquire a new color charge and the color force is mediated by eight massless gluons. QCD is unique in a number of aspects. It is a fully relativistic quantum field theory where the constituents are not accessible. The quarks and gluons are *confined*, i.e. only their color neutral combinations, i.e. hadrons, are detectable in experiments. As a consequence, most of the visible phenomena in QCD are emergent.

Further, the QCD coupling is *asymptotically free*, i.e. at high energies the coupling is relatively small and the theory is calculable. At low energies, the QCD coupling is relatively large and the theory at present is not calculable analytically. Numerical simulation of QCD using a discretized space-time lattice on the world's most powerful computers has been developed to the point where some important QCD observables can now be calculated *ab initio* with precision comparable to experimental uncertainty.

QCD emerged in the 1970s from the theoretical perspective of gauge symmetries and the experimental discovery that the proton was built from charged, pointlike constituents. Since that time, QCD has been validated in the regime where the coupling is small, i.e. at high energies. The comprehensive understanding of QCD as it applies at lower energies, i.e. in the energy regime where the coupling is large, remains elusive. In particular, understanding confinement is widely recognized as one of the major open questions in physics.

In a relativistic quantum field theory, the virtual particles play an integral role. In the quantum theory of electricity and magnetism (QED), the effects of the virtual particles are of order 1%. In QCD, the three *valence* quarks in the proton generate via interactions a large number of low-energy, virtual quarks, antiquarks and gluons. The effects of these virtual particles are dominant in describing the structure and properties of the proton. They were initially studied with lepton beams in the proton at the HERA collider but a comprehensive and precise investigation demands ion species from the proton to uranium, high luminosity, polarized beams, and optimized detectors.

For example, the *up* and *down* quarks have masses of order 5 MeV but the effects of their interactions via QCD result in a proton of mass 938 MeV. Thus, although the Higgs boson provides the 5 MeV quark mass, it is strong-coupling QCD that is responsible for essentially all of the mass of the luminous matter in the universe. Similarly, the origin of the spin-½ of the proton in terms of the constituent quarks and gluons has been studied at laboratories worldwide, including Jefferson Lab, but the understanding is far from complete. In high-energy experiments, the quarks and gluons scatter and, in a process known as *hadronization*, are detected in the form of hadrons. This process fundamentally connects the world of quarks and gluons, which is described by QCD, with the world of hadrons, where experiments live. We lack a fundamental understanding of this crucial connection.

---

[#]milner@mit.edu

One of the most significant achievements in physics over the last decade was the discovery at RHIC in ultra-relativistic, heavy-ion collisions of the hot, dense matter, which previously existed only at the dawn of the universe. It was found that this matter had the properties of a perfect fluid. Subsequently, this discovery was confirmed at the LHC and the new matter is being studied at both laboratories. However, careful measurements as a function of nuclear mass from the proton to uranium have raised some puzzles. Little is understood about how this matter is created and what its properties are. A solid understanding in terms of QCD will require a determination of the initial conditions of the heavy-ion collision as well as an understanding of hadronization.

The experimental study of QCD remains a scientific priority internationally with major facilities operating and new accelerators planned in Asia, Europe, and the U.S. In the U.S., the two world-class facilities, Jefferson Lab and RHIC, are international centers for the study of QCD. Jefferson Lab has an unique electron microscope that has carefully and precisely studied the low energy limits of the asymptotically free regime. With the increase in beam energy to 12 GeV, these studies can be dramatically enhanced in the valence quark region. In addition to discovering and studying the hot dense matter in heavy ion collisions, described above, RHIC is the world's first and only polarized proton collider. A first, direct determination of the contribution of the gluons to the proton's spin has been established.

Major advances in theory support a new era of precision QCD. The relentless drive for powerful supercomputers and more sophisticated algorithms ensure that the precision and applicability of lattice QCD will continue to become more potent. An effective field theory applicable to the high gluon density regime of hadrons, known as the *color glass condensate*, is widely used to interpret RHIC and LHC heavy ion data. Its study in lepton-nuclear collisions at high energy is a high scientific priority. Further, QCD theorists have developed a new, unifying framework with which to describe the 3D structure of hadrons based on Wigner distributions, which provides a powerful means to guide experiments.

**In summary, the complete understanding of QCD, in particular in the strong-coupling regime, demands a new era of precision measurement and *ab initio* calculations. For the first time, physicists are in the position to undertake the detailed study of a fully relativistic, quantum field theory, QCD, that we know describes the real world. A complete understanding of the Standard Model demands such a detailed study.**

## WHY AN ELECTRON-ION COLLIDER?

The world-wide QCD community has carefully studied for almost two decades [1-3] the optimized experimental configuration needed to carry out a precision study of QCD, as motivated above. This culminated in the statement in the 2007 U.S. Nuclear Physics Long Range Plan, *The Frontiers of Nuclear Science:* "An Electron-Ion Collider (EIC) with polarized beams has been embraced by the U.S. nuclear science community as embodying the vision for reaching the next QCD frontier" [4]. Electron beams couple to the charges in hadrons and are the established tool of choice to experimentally determine the constituent distributions at either the nucleon or fundamental quark and gluon level. Moreover, at high energies where the electron-quark scattering is dominant, experiments are analyzed in terms of snapshots of the constituents. These snapshots are directly interpretable in QCD. The energy and kinematics define the resolution ($Q^2$) and momentum scale ($x$) at which the snapshot is taken. A major thrust of EIC is to carry out three-dimensional tomography at the fundamental quark and gluon level of the structure of the nucleon. This raises the tantalizing possibility that a visualization of the microcosm can be realized for the first time.

The fact that the electron-quark coupling is relatively weak is advantageous in terms of theoretical interpretation. However, experimentally it has the very important consequence that the collision luminosity must be maximized. This is one of the key design parameters of the electron-ion collider.

The desire to reconstruct the final-state in high-energy electron-ion collisions naturally leads to a collider configuration. In addition, a large kinematic range (both $x$ and $Q^2$) enables study of the quarks and gluons in the proton and in nuclei. For example, these distributions constitute the initial state of ultra-relativistic heavy-ion collisions and at present are unmeasured experimentally.

The electron probe also allows variation of the energy transferred to the struck quark independently of the resolution $Q^2$. This is essential for understanding hadronization in QCD. The energy loss characterizes the length scale of the quark-to-hadron evolution.

Polarized electron and nucleon capabilities are another essential design parameter of EIC. They are required to probe the origin of the nucleon spin. In addition, polarization observables can provide sensitivity to important observables by careful choice of the kinematics. High precision polarimetry is key to maximally exploiting the polarized beams.

Finally, a suite of optimized detectors is required to carry out the necessary measurements. Ability to detect high rates, large acceptance, and clean particle identification are some of the demands that shape the design of these key scientific instruments.

Realization of EIC and its associated scientific instrumentation will require development of challenging, new technology. The U.S. flagship laboratories, BNL and Jefferson Lab, together with university researchers in collaboration with international partners and industry have the expertise and experience to deliver this in a timely way, once EIC is established as a national scientific priority.

## EIC SCIENCE HIGHLIGHTS

The development of the science case for the next QCD machine has been developed at many focused meetings involving experimenters, theorists, and accelerator physicists across the U.S., in Europe, and in Asia over the last two decades. Following the 2007 U.S. Long Range Planning Exercise, groups working on EIC detector and accelerator design were established at the major U.S. QCD laboratories, BNL and JLab. In addition, these laboratories put in place an EIC Advisory Committee comprising international expertise in theory, experiment, and accelerators. The EICAC has met regularly and provided valuable advice on EIC activities. In fall 2010, a ten-week workshop at the Institute for Nuclear Theory, University of Washington, Seattle gathered all interested physicists and assembled a broad compelling science case [5]. In 2013, a subgroup of leading QCD physicists appointed by BNL and JLab produced a succinct and compelling white paper: *Electron Ion Collider: The Next QCD Frontier – Understanding the glue that binds us all* [6]. Here, I select a few highlights from the white paper.

### The Nucleon Spin and its 3D Structure and Tomography

Several decades of experiments on high-energy lepton-quark scattering have provided an understanding of how the quarks and gluons share the longitudinal momentum of a nucleon. They have not, however, resolved the question of how the constituents give rise to the nucleon's spin, mass and magnetic moment. Unlike earlier studies, the EIC is designed to produce multi-dimensional maps of the distributions of the quarks and gluons in space, momentum, spin, and flavor.

For example, the dramatic kinematic reach and precision of EIC will have a huge impact on the quest to understand the origin of nucleon spin. This is made clear by Fig. 1.

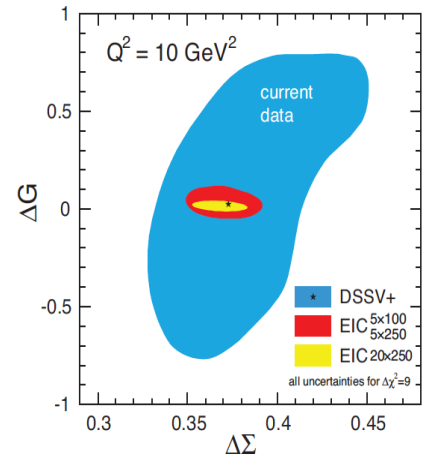

Figure 1: The projected reduction in the uncertainties of the gluons' contribution $\Delta G$ vs. the quark contribution $\Delta \Sigma$ that would be achieved by EIC for different center-of-mass energies.

Further, semi-inclusive measurements will provide a determination of the correlations between the motion of the quarks and their spins, as well as the spin of the parent nucleon. These correlations can arise from spin-orbit coupling of the quarks, about which very little is known at present. In this way, the full 3D dynamics of the proton can be studied, far beyond what is possible with conventional distributions. With both electron and nucleon beams polarized, EIC will allow physicists to study and understand the motion of the virtual QCD particles for the first time. For example, in Fig. 2 the color indicates the probability of finding the *up* quarks at a certain momentum in a transversely polarized proton.

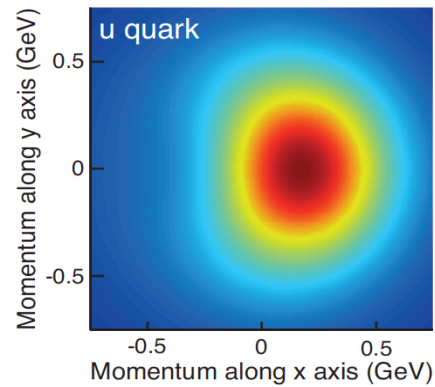

Figure 2: The transverse momentum distribution of an *up* quark with longitudinal momentum fraction $x = 0.1$ in a transversely polarized proton.

## The Nucleus, a QCD laboratory

Understanding the structure and properties of nuclei in QCD is a primary goal of nuclear physics. EIC will be the first experimental facility capable of exploring the internal 3D virtual quark and gluon structure of a nucleus at high-energy. Indeed, the nucleus in its ground state is an unprecedented and unexplored QCD laboratory for discovering and studying the predicted dense, gluonic matter as well as for studying the propagation of fast-moving color charges in the nuclear medium.

In QCD, the large soft-gluon density drives the non-linear gluon-gluon recombination. This gives rise to a saturation scale $Q_s$, at which gluon splitting and recombination reach a balance. At this scale, the density of gluons is expected to saturate, and may produce new and universal forms of hadronic matter. This is shown schematically in Fig. 3. The *color glass condensate* is a widely-used, theoretical framework for describing the saturated, soft gluonic matter. There are strong hints from HERA, RHIC, and LHC that the framework is valid, but lepton-nuclear experiments at high energy will be definitive.

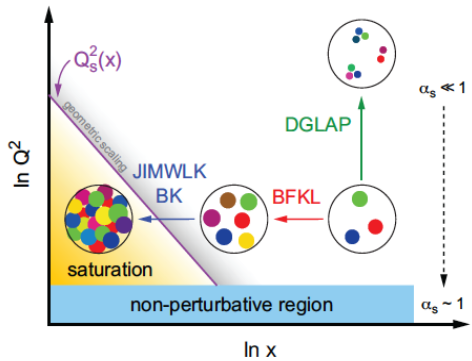

Figure 3: Resolution vs. energy indicating regions of weak and strong coupling QCD. In the latter the transition from low to highly-saturated quark and gluon density is indicated.

Fig. 4 illustrates some of the dramatic, predicted effects of gluon density saturation in electron-nucleus vs. electron-proton collisions at EIC. Gluon saturation greatly enhances the fraction of the total cross-section accounted for by diffractive events. As high luminosity is not necessary, an early measurement of coherent diffraction in e+A collisions at EIC would provide the first unambiguous evidence of gluon saturation.

The nucleus remains poorly explored by high-energy lepton scattering. We know since the mid 1980's that the quark momentum is modified in the nucleus but the distribution of the gluons is undetermined experimentally. These gluons are the dominant initial-state participants in the collision of ultra-relativistic heavy ions where the hot, dense matter is generated. EIC could obtain the spatial distribution of gluons in a nucleus by measuring the coherent diffractive production of J/ψ in e+A scattering.

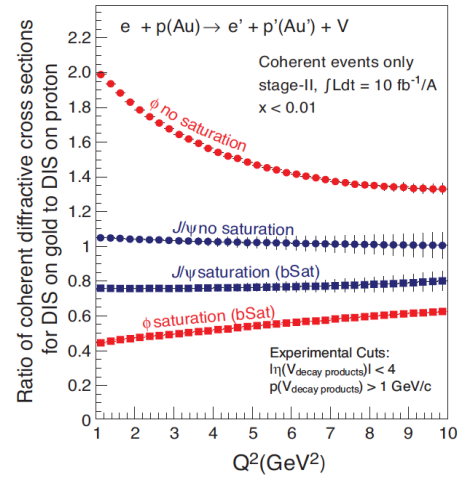

Figure 4: The ratio of a specific (coherent diffractive) cross section in $e+Au$ to $e+p$ collisions plotted vs. $Q^2$ for different models.

One of the key pieces of evidence for the discovery of the hot, dense matter at RHIC is jet quenching, manifested as a strong suppression of fast-moving hadrons. The suppression is believed to be due to the energy loss of quarks and gluons traversing the hot, dense matter. A puzzle is the fact that the production is nearly as much suppressed for heavy as for light mesons, since a heavy quark is much less likely to lose its energy via medium-induced radiation of gluons. The variety of ion beams available for e+A collisions at EIC would provide a femtometer filter to test and to determine the correct mechanism by which quarks and gluons lose energy and hadronize in nuclear matter, as shown in Fig. 5.

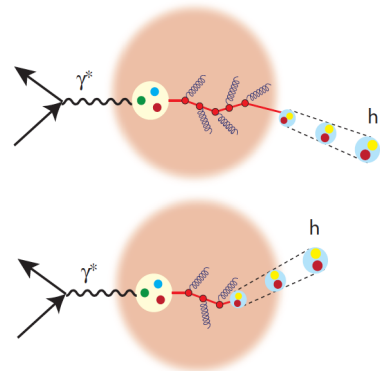

Figure 5: Schematic figure of a quark or gluon moving through cold nuclear matter: the hadron can be formed either outside (top) or inside (bottom) the nucleus.

*Beyond the Standard Model*

With the new combination of experimental probes, high center-of-mass energy, high luminosity, and the ability to polarize the electron and hadron beams, EIC offers significant opportunities to search for new physics beyond the Standard Model. This can come through the ability to detect rare reactions with increased sensitivity or to carry out measurement of fundamental SM parameters with unprecedented precision.

For example, a search for charged lepton flavor violation in e -> τ transition looks very promising. The collider environment event topology for rare signal events can be differentiated from conventional electroweak deep-inelastic scattering events.

In the Standard Model, weak neutral current couplings are all functions of the weak mixing angle $\sin^2\theta_W$. Precision measurement of the weak mixing angle is an established means to look for new physics. The parity-violating asymmetry in spin-dependent electron-proton and electron-deuteron scattering can be measured at EIC in the momentum transfer region where the weak mixing angle is changing most rapidly. Fig. 6. shows a preliminary projection of statistical uncertainties for a center-of-mass energy of 140 GeV and an integrated luminosity of 200 fb$^{-1}$.

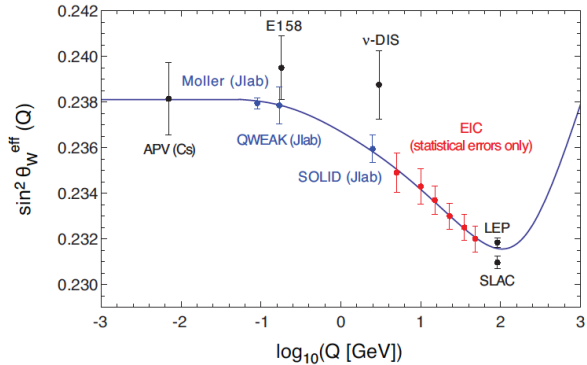

Fig.6. Projected statistical uncertainties in measurement of $\sin^2\theta_W$.

## THE PATH FORWARD

The U.S. nuclear physics community will develop a new Long Range Plan in 2015 with a perspective on strong interaction frontier research looking out to about 2025. In that timeframe, it is expected that the JLab 12 GeV upgrade will come into routine operation, the FRIB accelerator will come online and begin operation, and the RHIC accelerator, as presently configured, will come to the end of it scientific life. The U.S. QCD community will make the case that a new era of precision study is demanded to complete the understanding of the fundamental structure of matter. The central component of such a program will be a new, high-luminosity, polarized electron-ion collider. If EIC is established as a national scientific priority, it can be realized within about a decade. Such a capability is essential to maintain U.S. leadership in a central area of fundamental, subatomic physics research.

## ACKNOWLEDGMENT

The author's research is supported by the DOE Office of Nuclear Physics. The EIC science case is the result of dedicated effort by many experimentalists, theorists, and beam physicists over an extended period.